\documentstyle[12pt]{article}
\begin{document}

\pagestyle{empty} 

\begin{center}
{\Huge \bf STRINGS $'99$},
\end{center}
\vspace{1cm}
\begin{flushright}
Potsdam, July 19 - 24
\end{flushright}
\vspace{1cm}
\begin{center}
     {\Huge \bf Christos Epameinonda Kokorelis}\\[0pt]
\end{center}
\vspace{2cm}
\begin{center}
{\huge F-Theories with Reduced Monodromy }
\end{center}
\begin{center}
{\huge and}
\end{center}
\begin{center}
{\huge Effective String Theories}
\end{center}

\begin{center}
{\underline {\huge Abstract}}
\end{center}
We discuss aspects of the non-standard version of F-theory based on the
arithmetic of torsion points on elliptic curves. We construct new F-theory
vacua in 8-dimensions. They are coming by the projective realizations
of F-theory on K$_3$ surfaces admitting double covers onto $P^2$,
branched along a plane sextic curve, the so called double sextics.
The new vacua are associated
with singular K$_3$ surfaces. In this way the stable picture of the
heterotic string is mapped at the triple ponts of the sextic.
We argue that this formulation
incorporates naturally the $Sp(4,Z)$ invariance
that the extrapolating four dimensional vector multiplet sector
of all heterotic vacua may possess.
In addition, we describe the way that the 4${\cal D}$ genus two description
of (0,2) moduli dependece of $N=1$ gauge coupling constants
may be connected to Riemann surfaces, with natural $Sp(4,Z)$ duality
invariance.
Here, we recover a novel way to break space-time supersymmetry
and fix the moduli parameters  in the presence of the Wilson lines.
We also consider a novel way to break space-time supersymmetry
and fix the moduli parameters in the presence of the Wilson lines.
We also consider the heterotic duals to compactifications of F-theory
to four dimensions belonging to isomorphic classes of eliptic curves
with point cusps of order two. For the latter theories
we calculate the $N=2$ 4${\cal D}$ heterotic prepotential corresponding
to ${\Gamma_o(2)}_T \times {\Gamma_o(2)}_U$ classical
perturbative duality group and their conjugate modular theories.
\newpage

\begin{center}
{\underline {\huge Introduction}}
\end{center}
 Higher dimensional theories like F-theory don't have any
obvious manifestation in twelve dimensions as one should expect.
Instead F-theory makes its presence manifest only through its
compactifications to lower dimensions that match already known
compactifications of the heterotic string ( or its equivalent
theories IIA, IIB, etc). In turn $N=2$ 4$\cal D$ compactifications
of the heterotic string, an obvious candidate
for F-theory compactification match have a $N=2$ vector multiple sector
 defined by a K\"ahler potential
\begin{equation}
K=-\log (-i \Omega^{\dagger} \left( \begin{array}{cc}
0&1\\-1&0
\end{array}\right)\Omega)= -\log(i {\bar X}^I F_I - iX^I {\bar F}_I),
\;I=0,\dots,n
\label{rad1}
\end{equation}
 where  $\Omega=( X^I(M^I), F_I(X^I))$ the holomorphic
symplectic
period vector, and $X^I$ the special coordinates.
Demanding complete gauge fixing, in perturbation theory, of the
vector multiplet sector of the four dimensional $N=2$ compactifications
of the heterotic string requires the knowledge of the holomorphic
prepotential $F$. That has been calculated in
\cite{anafo1, anafo2, anafo3}.
\newline
 The full target space duality transformations act on the space
of the symplectic vectors $\Omega$, which includes Wilson lines,
in the form 
\begin{equation}
K= -\log[( T + {\bar T})(U + {\bar U}) - (B+ {\bar C})({\bar B} + C)]
\label{rad2}
\end{equation}
\begin{equation}
M= \left( \begin{array}{cc}T&B\\-C&U \end{array}\right),\;M\;\rightarrow
\;\left(\begin{array}{cc}a&b\\c&d\end{array}\right),\;
\left(\begin{array}{cc}a&b\\c&d\end{array}\right)\;\in\;Sp(4,Z).
\label{rad3}
\end{equation}

An obvious challenge, for F-theory, that is matching the existing
heterotic compactifications, may be directly related to the
arithmetic of torsion points on elliptic curves.
\newline
In this work, we will give evidence for the presence of new
8-dimensional F-theory realizations \cite{anafo5}.

\begin{center}
{\underline {\huge Double Covers}}
\end{center}
 It is known that F-theory compactified on a K$_3$ surface
realized as an elliptic fibration with a section
is on the same moduli space, namely
\begin{equation}
O(\Gamma_{2,18})\O(2,18)/(O(2) \times O(18))
\label{rad4}
\end{equation}
 the perturbative heterotic string on a $T^2$ torus.
Schematically,
\begin{equation}
\frac{F}{K_3} \equiv \frac{{(E_8 \times E_8)}_{het}}{T^2}.
\label{rad5}
\end{equation}
On the right hand side of this equivalence the torus
can be represented as an elliptic curve,
\begin{equation}
y^2 =x^3 + ax+ b,
\label{rad6}
\end{equation}
 that is the double cover of the complex plane x.
However, on the left hand side of (\ref{rad5}) the K$_3$ surface $S_F$ building
the compactification of F-theory is represented as an elliptic fibration
over a $P^1$ base,
namely as the map $\pi: S_F \rightarrow P^1$.
Comparison of the two sides in (\ref{rad5}),reveals the lack of the
presence of a double cover on the F-theory side. This constitutes
a form of {\bf violation of double cover parity}
signalling different treatment of the bases of the F-theory/heterotic duality
pairs.
 By demanding that F-theory respects the double cover treatment of its base
we may generate new F-theories. The double cover parity
may disappear only when we consider generating surfaces from {\bf double
covers onto $P^2$}.
Double covers of K$_3$ onto $P^2$, the so-called {\bf double sextics} are
branched
along a plane sextic curve. In general a curve X which is a double cover of
Y branched along  a
curve C, is one to one correspondence with singular K$_3$ surfaces e.g
maximal Picard number equal to twenty. Every singular surface is the
double cover of a K$_3$ surface.
For example, at the limit that the $T^2$ is large
the K$_3$ surface degenerates into a variety that is made from
 the union of two intersecting rational elliptic surfaces $S_1$, $S_2$
 intersecting along an elliptic curve $E^{*}$.

\begin{center}
\begin{tabular}{|c|c|r|} \hline
Cubic fibration & order F &
{$K_3$ fibration}\\ \hline\hline
$II^{*}$ $2I_1$&1& $2II^{*}$ $2I_2$\\ \hline
$II^{*}$ II&1& $2II^{*}$ IV\\ \hline
$III^{*}$ $I_2$ $I_1$&2& $2III^{*}$ $I_4$ $I_2$ \\ \hline
$III^{*}$ III&2&$2III^{*}$ $I_0^{*}$\\ \hline
$I_4^{*}$ $2I_1$&2&$2I_4^{*}$ $2I_2$\\ \hline
$I_9$ $3I_1$ &3&$I_{18}$ $I_2$ 4$I_1$, 2$I_9$ 2$I_2$ 2$I_1$ \\ \hline
$IV^{*}$ $I_3$ $I_1$&3&$2IV^{*}$ $I_6$ $I_2$\\ \hline
$IV^{*}$ IV&3&3$IV^{*}$\\ \hline
$I_8$ $I_2$ 2$I_1$ &4&$I_{16}$ $I_4$ 4$I_1$, $I_{16}$ 3$I_2$ 2$I_1$,
$I_{16}$ 3$I_2$ 2$I_1$, 2$I_8$ $I_4$ $I_2$ 2$I_1$,
2$I_8$ 4$I_2$\\ \hline
$I_2^{*}$ $2I_2$&4&$2I_2^{*}$ $2I_4$\\ \hline
$I_1^{*}$ $I_4$ $I_1$&4& $2I_1$ $I_8$ $I_2$\\ \hline
2$I_5$ $2I_1$&5&2$I_{10}$ $4 I_1$, $I_{10}$ $2I_5$ $I_2$ $2I_1$,
$4I_5$ $2I_2$\\ \hline
$I_6$ $I_3$ $I_2$ $I_1$&6&$I_{12}$ $I_6$ $2I_2$ $2I_1$, $I_{12}$ $I_4$
$2I_3$ $2I_1$, $I_{12}$ $2I_3$ $3I_2$, $3I_6$ $I_4$  $2I_1$,\\ 
&&$3I_6$ $3I_2$, $2I_6$ $I_4$ $2I_3$ $I_2$\\ \hline
$2I_4$ $2I_2$&8& $2I_8$ $4I_2$, $I_8$ $3I_4$ $2I_2$, $6I_4$ \\ \hline
$4I_3$&9&$2I_6$ $I_3$\\ \hline 
\end{tabular}
\end{center}
\begin{center}
Table 1
\end{center}

 All the possible forms of the rational elliptic surface as an
 elliptic fibration against its double cover and the order of the
 Mordell-Weyl (MW)group are listed in table 1. 
The first column 
gives us the the Kodaira fibers appearing in the cubic associated with
 the corresponding Mordell-Weyl group order of the group of sections
given in the second column. The third column is the configuration of
allowed Kodaira fibers for the associated $K_3$ fibrations.

The double cover
 consists of an elliptic fibration $\pi : Y \rightarrow P^1$ and an
 involution $\sigma$that respects the fibration . In this case the
 fibration
  always have fixed points.
  \newline
  The form of the rational elliptic fibration is directly
  connected to the order of MW that defines a group operation
  over the rational points Q of the elliptic curve.
  Its action restricts the form of the associated Weierstrass
  fibration and cam be written
   as $E(Q) \equiv Z^R \oplus \Phi$, where r is the rank of
    $E(Q)$ and $\Phi$ the torsion subgroup.
    So for example for the
    \newline
    \begin{equation}
    \bullet\; \Phi \cong Z_2 \oplus Z_2,\;\;\;\;\;\;\;\;\;\;\;\;\;
    \;\;\;\;\;\;\;\;\;\;\;\;\;\;\;\;\;\;\;\;\;\;\;\;\;\;\;
    \;\;\;\; y^2 = x(x-\beta)(x-\gamma),
    \label{rad7}
    \end{equation}
     while for the
    \newline
    \begin{equation}
    \bullet\; \Phi \cong Z_4 \oplus Z_2,\;\;\;\;\;\;\;\;\;\;\;\;\;\;
    \;\;\;\;\;\;\;\; y^2 =
    x(z+ \zeta^2)(x+ \lambda^2),\;\beta=\zeta^2,\;\gamma=\lambda^2.
    \label{rad8}
    \end{equation}
  \begin{center}
 {\underline {\huge Effective String Theories from Double Covers}}
\end{center}

On the previous section we demanded that the base space of the
8-dimensional F-theory/ heterotic duality pair must be treated
in terms of double covers.
The question that remains now is that if we can find an effective
theory that can be formulated in terms of double covers and is defined
in lower dimensions , e.g four, that may be defined both
in the F-theory and the heterotic side. The case of the heterotic string
is considered in this section while the $N=2$ four dimensional
compactification of F-theory counterpart will be treated in a future work.

The hyperelliptic curve for a genus 2g+2 resp. 2 surface 
is given by
\begin{equation}
y^2=\Pi_{i=1}^{2g+2}(x-e_i)=P_{2g+2}(x, e_i),\;e_i\; \neq\; e_j,for\;i \neq j,
\label{rad9}
\end{equation}
  and represents the double cover of the sphere 
branched over 2g+2 resp. 2 points.
  The map from the Jacobian to the complex numbers defines the $\Theta$
  functions with the usual build $Sp(2g, 2)$ invariance. So in $g=2$
  the invariance is $Sp(4,Z)$ the required T-duality of the
  perturbative heterotic string spectrum.
  {\em In order to define the genus 2 hyperelliptic curve
  as the Riemann
  surface that the (0,2) 4$\cal D$ heterotic string lives} we need
  one more element.
  The element that we require is that {\em there is a birational
  correspondence \cite{anafo4} between the projective varieties}
  associated with the graded ring of even projective invariants
  of binary sextics and with the graded ring 
of modular forms. In simple terms that means that the projective variety
linked with the graded ring of even projective invariants of binary
sextics
is a compactification of moduli of genus
  two. Therefore if we denote the six roots $\phi_I,\;I=1,\dots,6$ of 
    the sextic
  $s_o X^6+ s_1 X^5+\dots + s_6$  and we denote their difference
  by ($\phi_i -\phi_j$) the invariants A, B, C, D take the
  following form

\begin{eqnarray}
A(s)=s_o^2 \sum_{fifteen}(12)^2 (34)^2 (56)^2\nonumber\\
B(s)=s_o^4 \sum_{ten}(12)^2 (23)^2(31)^2(45)^2(56)^2(64)^2\nonumber\\
C(s)=s_o^6\sum (12)^2(23)^2(31)^2(45)^2(56)^2(64)^2(14)^2(25)^2(36)^2
\nonumber\\
D(s)=s_o^{10}\Pi_{i<j}(jk)^2.
\label{rad10}
\end{eqnarray}
   Because a sextic can be brought into the general form
\begin{equation}
X(X-1)(X-\lambda_1)(X-\lambda_2)(X-\lambda_3)
\label{rad11}
\end{equation}
   we can replace each of the three lambda in (\ref{rad11}) by some
  theta function of zero argument, namely
\begin{equation}
\lambda_1 =(\frac{\theta_{1100}\theta_{1000}}{\theta_{0100}\theta_{0000}})^2,\;
\lambda_2=(\frac{\theta_{1001}\theta_{1100}}{\theta_{0001}\theta_{0100}})^2,\;
\lambda_3=(\frac{\theta_{1001}\theta_{1000}}{\theta_{0001}\theta_{0000}})^2,
\label{rad12}
\end{equation}
 where
\begin{equation}
\theta_{g_1 g_2 h_1 h_2}\left(\begin{array}{cc}\tau_1&\epsilon\\
\epsilon&\tau_2\end{array}\right)=\sum_{n=0}^{infty}
\frac{2^{2n}}{(2n)!}\frac{d^n}{(d\tau_1)^n}\theta_{g_1 h_1}(\tau_1)
\frac{d^n}{(d\tau_2)^n} \theta_{g_2 h_2}(\tau_2) \epsilon^{2n}
\label{rad13}
\end{equation}
   theta functions of genus two. It can be proved that $\lambda_1, \lambda_2, 
\lambda_3$
  can be expanded in terms of even powers of $\epsilon$ when $\epsilon$ is
small.
      As a result all the variables in (\ref{rad11}) are fixed and 
its roots may be
calculated.
Moreover the invariants A, B, C, D, are expressed in terms of $\epsilon$ 
and    .
the following relations hold
\begin{eqnarray}
I_4=\sum(\theta_m)^8,\nonumber\\
I_6=\sum \sum \pm (\theta_{m_1} \theta_{m_2} \theta_{m_3})^4,\nonumber\\
I_{10}= -2^{14} \cdot \chi_{10}= \Pi (\theta_m)^2,\nonumber\\
I_{12}=2^{17} 3 \cdot \chi_{12} = \sum(\theta_{m_1} \theta_{m_2}\dots\theta_{m_6})^4,
nonumber\\
I_{35}=2^{39} 5^3 i \cdot \chi_{35}  = (\Pi \theta_m)(\sum_{azygous}
\pm (\theta_{m_1} \theta_{m_2} \theta_{m_3} )^{20}).
\label{rad14}
\end{eqnarray}
It is remarkable that the following relations hold
\begin{equation}
D/A^5 \propto \epsilon^{12},(B/A^2)^3 \propto j(\tau_1) j(\tau_2)
\epsilon^{12}, ((3C-AB)/A^3)^2 \propto (j(\tau_1-j(\tau_2))\epsilon^{12}.
\label{rad15}
\end{equation} 
  or in precise form
\begin{equation}
I_4 =B,\;I_6 =(\frac{1}{2}(AB-3C),\;I_{10}=D,\;I_{12}=AD,\;I_{35}5^3 D^2 E.
\label{rad16}
\end{equation}
That means that the uniformation parameters of the sextic (\ref{rad11}) are 
fixed
genus two elements therefore possessing manifest $Sp(4,Z)$ invariance.
In its non-perturbative form the equation for the sextic $P_6$ involves
the parametric relation (\ref{rad16}). Notice that in (\ref{rad15}) the
fundamental invariant of the full theory are expressed in terms of products
of j-invariants. What we have not discuss is the expansion of genus two theta
functions in terms of epsilon parameter, that coincide with the Wilson
line background fields, represents exactly the fact that the space
of projective varieties corresponding to the invariants A, B, C, D has been
blown up such that the jacobian variery of the genus two curve has
degenerate to a product of elliptic curves. The blow up process is necessary
since the projective variety $P_6$ does not apriori include the Siegel
fundamental domain. The correspondence with the heterotic string
comes after identifying $T=\tau_1$, $U= \tau_2$, $\epsilon =$Wilson line.
\newline
At the points that the discriminant of the projective variety
of $P_6$ degenerates both T, U and the Wilson line are involved in a
non-trivial relation. That means that one or more moduli may be fixed
therefore breaking space-time supersymmetry.

 At this point we might be tempted to give a description
 of the $N=2$ vector multiplet
 effective theory in the case that effective heterotic theory
 comes with target space duality group ${\Gamma_o(2)}_T \times
 {\Gamma_o(2)}_U$. The one loop perturbative heterotic prepotential
 can be calculated in this case \cite{anafo5}.

 \begin{center}
 {\underline {\huge Conclusions}}
 \end{center}
  In the previous sections {\em we introduced a new constraint},
 that the F-theory/heterotic duality may satisfy in order
 to correctly reproduce the target space duality
 transformations of the heterotic string.
 That involves the treatment of the base in both pairs
 in terms of double covers.
 \newline
 Particular role in our study is layed by the MW group.
 Its presence generates new solutions for the stable degeneration
 limit of the F-theory K$_3$ surface.
 \newline
 At the moment our study is confined on the equivalence of the
 F-theory/heterotic string at 8-dimensions in terms of double covers.
 The relevance of our results for the $N=2$ 4$\cal D$
 theory comes only through
 the determination of the Riemann surface that is a double cover
 of the sphere. In order for our results to be relevant to the
 F-theory/heterotic duality in four dimensions we must find
 the F-theory compactification realized on a four
 dimensional context.
 Work is in progress towards this goal.


\begin{thebibliography}{500}
\bibitem{anafo1} B. de Wit, V. Kaplunovsky, J. Louis and D. Lust,
Nucl. Phys. B451 (1995) 53, hep-th/9504004.
\bibitem{anafo2} I. Antoniadis, S. Ferrara, E. Gava, K.S.Narain and
T.R.Taylor, Nucl. Phys. B447 (1995) 35, hep-th/9504034.
\bibitem{anafo3} C. Kokorelis, Nucl. Phys. B542 (1999) 89-111.
\bibitem{anafo4} J. Igusa, Amer. Jour. of Math. Vol. 89 (1967) 817.
\bibitem{anafo5}C. Kokorelis, F-Theories on Double Sextics and Effective
String Theories hep-th/9901150.
\end{thebibliography}
\end{document}